\documentclass{article}
\input amssym.def
\input amssym.tex

\def\ben{\begin{equation}}
\def\een{\end{equation}}

\def\bea{\begin{eqnarray}}
\def\eea{\end{eqnarray}}

\begin{document}

\hfuzz=100pt
\title{The Ashtekar-Hansen universal structure at spatial infinity
 is  weakly pseudo-Carrollian.  
}
\author{G W Gibbons 
\\
D.A.M.T.P.,
\\ Cambridge University,
\\ Wilberforce Road,
\\ Cambridge CB3 0WA,
 \\ U.K.
}
\maketitle

\begin{abstract} 
  It is shown that Ashtekar and Hansens's
  Universal Structure at Spatial Infinity
  (SPI), which has recently be used to establish the conservation of
 supercharges from past null infinity to future null infinity,
 is an example of a (pseudo-) Carrollian structure.
The relation to  Kinematic Algebras is clarified.

\end{abstract}

\section{Introduction}

In a recent paper \cite{Prabhu:2019fsp} 
it has been argued that the conservation of supercharges
from past to future null infinity holds provided the spacetime is
asymptotically flat in the sense of The Ashtekar and Hansen
\cite{Ashtekar:1978zz}. The purpose of the present paper is to show that t
he universal structure at infinity that such spacetimes admit is an example
of what has been defined as a Carrollian structure
\cite{Duval:2014uoa,Duval:2014uva}  under the novel
assumption that the degenerate  Carrollian metric is pseudo-Riemannian
rather than Riemannian. The relation of this structure to  older
\cite{Bacry:1968zf,Bacry:1986pm,Nzotungicimpaye:2014wya} 
and more recent
\cite{Figueroa-OFarrill:2017sfs,Figueroa-OFarrill:2017ycu,Figueroa-OFarrill:2017tcy,Andrzejewski:2018gmz,Figueroa-OFarrill:2018ilb,Figueroa-OFarrill:2018ygf}
work on Kinematic Algebras is described.

\section{The Ashtekar-Hansen  universal structure on Spi.} 

In \cite{Ashtekar:1978zz} the authors constructed a 4-manifold
Spi with a {\it universal structure} associated to the spatial
infinity, $i^0$   for spacetimes which are
{\it asymptotically empty and flat at null  and spatial infinity }
(AEFANSI).  Spi  should be construed
as a a blow up of the point  $i^0$  in the conformal embedding of the
physical spacetime which captures
the idea that  $i^0$ is the ultimate destination of spatial
curves. The equivalences class of  curves is specified by
an ultimate direction and a notion of
ultimate acceleration. 

The construction was later revisited
in terms of  a timelike boundary of an   
asymptotically flat  spacetime whose normal is only defined up
to a direction \cite{Ashtekar:1991vb}.

The four-manifold ${\rm Spi} =\{ E, \pi,  B\}$ is a
principal line bundle $E$  over
the set $B$ of unit spacelike four vectors in the tangent space of $i^0$ 
 with structural group $\Bbb{R}$.
 The base space $B$ of the line bundle Spi may be identified
 with three dimensional  de-Sitter
 spacetime $dS_3$ equipped with its Lorentzian metric $g_3$  of
 constant curvature.
Using the projection map $\pi: {\rm Spi} \rightarrow B= dS_3$ , the Lorentzian
 metric $g_3$
 may be pulled back to the 4-manifold Spi to give a degenerate bilinear form
  $g_4=  \pi^\star g_3 $ on the tangent space of  Spi
 with kernel tangent vectors
 to the fibres.

 The automorphisms of this structure is an infinite dimensional group $G$ 
 analogous to the BMS groups of future and past null infinity ${\cal I}^{\pm}$.  

 The group of such diffeomorphism
 obviously includes the  Lorentz group $SO(3,1)$ acting on the base $B$ .
 In addition, since there is no natural coordinate on the
 fibres of the bundle, that is no natural
 section, the space of all sections, i.e. functions
 on the base space,   form an infinite dimensional  abelian  subgroup group
 under addition called {\it {\rm Spi} supertranslations} which we call $T$ .
In fact
\ben
G= SO(3,1) \ltimes T \,, \qquad G/T= SO(3,1)  \,. 
\een
Moreover there is a 4 dimensional normal subgroup of translations $T_4$.

\section{The Weakly Carrollian structures } 
The universal  structure on Spi described above,
especially its degenerate metric
closely resembles what has been called a
Carrollian structure \cite{Duval:2014uoa,Duval:2014uva} and it has been shown how  the BMS group
acts  the automorphism  group of the Carrollian structure
on ${\cal I}$ which is a principal  line bundle over $S^2$ with fibre
$\Bbb R$  and a degenerate bilinear form whose kernel consists
of tangent vectors to the fibres.  Apart from its dimension
the main difference is that the metric on the base space
of the Carrollian structures considered previously was positive definite.
However the general ideas go through of one merely requires that
the metric on the base  is non-degenerate, i.e. pseudo-Riemannian.
It seems reasonable to refer to such structures 
as {\it pseudo-Carrollian }.  In \cite{Duval:2014uoa}  by analogy with a Newton-Cartan structure,
a strong Carrollian structure which included an affine connection was defined. But in  \cite{Duval:2014uva}  weaker
definition was adopted and it is this weaker definition which seems to be more appropriate in the present case.

Thus a pseudo-Carrollian  manifold may be defined as  a triple $(C,g, \xi )$
where $C$ is a smooth (d+1) dimensional twice covariant
degenerate symmetric tensor field $g$ whose kernel is generated by the
nowhere vanishing, complete vector field $\xi $.
The associated Carroll group $Carr(C,g ,\xi)$   are its   automorphisms
and  consist of
all diffeomorphisms
of $C$ preserving the bilinear form $g$ and the vector field $\xi$.
The {\it pseudo-Carroll Lie algebra}, $\frak{carr}(C,g ,\xi)$
is then identified with the Lie of those vector fields $X$ on $C$
such that

\ben
L_Xg =0\,, \qquad L_X \xi =0\,. \label{Lie}
\een

In fact (\ref{Lie}) are the same equations which were used in \S 4 of
\cite{Ashtekar:1978zz} 
to determine the asymptotic symmetries at infinity of AEFANSI spacetimes admitting a Spi.

There is an associated {\it Conformal Carroll group of level N},
$CCarr_N(C,g, \xi)$ whose transformations preserve  the tensor field $g \otimes \xi^{\otimes N}$
canonically associated with a Carroll manifold, that is all
diffeomorphisms $f$  satisfying
\ben
f^\star g = \Omega ^2 g \,, \qquad f^\star \xi = \Omega ^{-\frac{2}{N} } \xi 
\een
for some positive function $\Omega$ and integer $N$. The Lie algebra
of infinitesimal conformal Carroll transformations  
$\frak{ccarr}_n(C,g, \xi) $ is  spanned by
vector fields $X$ such that
\ben
L_Xg = \omega g \,, \qquad L_X \xi = -\frac{\omega }{N} \xi \,,
\een
for some function $\omega$ on $C$ \footnote{factor of 2 missing ?}.

\section{Kinematical algebras and Kinematical Spacetimes}
In \cite{Bacry:1968zf} (see also
\cite{Bacry:1986pm,Nzotungicimpaye:2014wya,Figueroa-OFarrill:2017sfs,Figueroa-OFarrill:2017ycu}.) Levy-Leblond
and Bacry introduced the idea  of a {\it kinematical algebra},  a
10-dimensional algebra $ \frak{k} = {\bf J} \oplus H \oplus {\bf
  P}\oplus {\bf B}$ where ${\bf J} = \frak{so(3)}$, $H= \Bbb{R}$,    $
{\bf P} =\Bbb{R}^3$ , $ {\bf B}= \Bbb{R}^3$ are rotations, time
translations,  space translations and boosts respectively,
and is a deformation of the so-called {\it Static kinematic algebra} for which
the only non-vanishing brackets are
\ben [{\bf J}, {\bf J} ]  \in {\bf
  J} \,,\qquad [H,{\bf J}]=0\,,\qquad [{\bf J}, {\bf P}] =\in {\bf
  P}\,,\qquad  [{\bf J}, {\bf B}] =\in {\bf B}\,.
\een

Bacry and Levy-Leblond
\cite{Bacry:1968zf} found 12 such algebras (or 11 if one insists that
the boosts ${\bf B}$ be non-compact generators).  All may be
obtained by contractions starting either from  $\frak{so}(3,2) $ or
$\frak{so}(4,1)$, both of which contract to the Poincar\'e algebra
$\frak{p}$ in the limit that the curvature goes to zero.  Performing a
further contraction one finds that as the velocity light goes to
infinity one obtains the Galilei group $\frak{g}$. On the other hand
as it goes to zero one obtains the  finite, 10 dimensional, Carroll
subalgebra algebra $\frak{c}$ of the infinite dimensional Carroll  algebra
$\frak{Carr}(3,1)$.  Starting from  $\frak{so}(3,2) $ one obtains
an algebra isomorphic to the the Poincar\'e algebra $\frak{p}$, called
by Bacry and Levy-Leblond 
the Para-Poincar\'e algebra $\frak{p}^\prime$ in which the boosts ${\bf B}$
and translations ${\bf P}$ are interchanged.

Each kinematical algebra $\frak{k}$ gives rise to a kinematical group $K$.
Figueroa-O'Farrill and Prohazka
have shown
\cite{Figueroa-OFarrill:2018ilb}that
associated with each such kinematical group are  one or more
{\it Kinematical Spacetimes}, that is  
4 dimensional  homogeneous  spacetimes
$M = K/H$ where the group $H$ has lie algebra
$\frak{h}$ and the pair $\frak{k},\frak{h}$ are subject to
certain admissibility conditions. 

In the case of the para-Poincar\'e  Group   $P^\prime$, which
\cite{Figueroa-OFarrill:2018ilb} call $AdSC$, the structure induced
on the associated Kinematic spacetime is Carrollian.
If we write the Anti-De-Sitter metric as
\ben
ds ^2 =-c^2(1+\frac{r^2}{R^2} ) dt ^2 + \frac{dr^2}{(1+\frac{r^2}{R^2} )}
+ r^2 (d \theta ^2 + \sin^2 d\phi^2 )  
\een
and take the limit $c^2 \downarrow  0$ we obtain
a Carrollian metric with $g_{ij}$ the metric on hyperbolic three space
${\Bbb H}^3$.

In fact it was shown in   \cite{Duval:2014uoa,Duval:2014uva}  that one may
obtain the Carroll structures as the data induced on a null
hypersurface in a $d+2$ pseudo Riemannian manifold
endowed with a Bargmann structure, in  particular on the  null hyperplane
$x^+ = {\rm constant}$  from
five-dimensional Minkowski spacetime with metric.   
\ben
ds ^2 = 2 dx^+ dx^- + d {\bf x}^2 \,.
\een

In  \cite{Figueroa-OFarrill:2018ilb} this procedure was  
adapted to obtain this  Carrollian  structure by considering
5-dimensional anti-de-Sitter spacetime $AdS_5$  with metric. 
\ben
ds ^2 = \frac{1}{z^2} \{ 2 dx^+ dx ^- + dz^2  +  (d x^1)^2 + (dx^2)    \}\,. 
\een

A null hypersurface is obtained by setting  $x^^+ = {\rm constant}$
and the degenerate bilinear  form  is
\ben
ds ^2 = \frac{1}{z^2} \{  dz^2  +  (d x^1)^2 + (dx^2)    \}\,, 
\een
which is the upper half space model of hyperbolic three space
${\Bbb H}^3$. By thinking of the $AdS_5$ as a quadric in ${\Bbb E} ^{4,2}$
one sees that action of $AdSC$ is clearly the subgroup of
$SO(4,2)$  leaving invariant  the intersection with the hypersurface
$ x^+ = {\rm constant}$. 

\section{Lorentzian Kinematic Algebras and Kinematical spacetimes}

It is clear that at the formal level that  much of the previous section will
go through with $ \frak{so}(3) $ replaced  by $\frak{so}(2,1)$.
The analogue of the spacetime associated to $AdSC$ is Spi.

\section{Acknowledgement}. The author thanks Gary Horowitz and Jorge Santos
for an enquiry and subsequent  discussions about 
geometric structures  at spatial infinity which stimulated my initial
interest  on some of the material described in this  paper.

\end{document}